# Field effect and photoconduction in Au$_{25}$ Nanoclusters Films

*Michael Galchenko, Andrés Black, Leonard Heymann, and Christian Klinke\**

M. Galchenko,[+] Dr. A. Black,[+] L. Heymann, Prof. C. Klinke
Institute of Physical Chemistry, University of Hamburg, Martin-Luther-King Platz 6, 20146 Hamburg, Germany
E-mail: christian.klinke@swansea.ac.uk
Prof. C. Klinke
Department of Chemistry, Swansea University – Singleton Park, Swansea SA2 8PP, United Kingdom
Prof. C. Klinke
Institute of Physics, University of Rostock, 18059 Rostock, Germany

[+]These authors contributed equally to this work

Keywords: Au nanoclusters, semiconductors, colloid chemistry, field effect, photoconduction

Quantum confined Au nanoclusters exhibit molecule-like properties, including atomic precision and discrete energy levels. The electrical conductivity of Au nanocluster films can vary by several orders of magnitude, and is determined by the strength of the electronic coupling between the individual nanoclusters in the film. Similar to quantum confined, semiconducting quantum dots, the electrical coupling in films is dependent on the size and structure of the Au core and the length and conjugation of the organic ligands surrounding it. Unlike quantum dots, however, semiconducting transport has not been reported in Au nanocluster films. We demonstrate that through a simple yet careful choice of cluster size and organic ligands, stable Au nanocluster films can electronically couple and become semiconducting, exhibiting electric field effect and photoconductivity. The molecule-like nature of the Au nanoclusters is evidenced by a hopping transport mechanism reminiscent of doped, disordered organic semiconductor films. These results demonstrate the potential of metal nanoclusters as a solution processed material for semiconducting devices.

Individual organic semiconductor (OSC)[1] molecules and colloidal semiconducting quantum dots (SQDs)[2] possess discrete energy levels. Collective properties can emerge in films of these materials if they are electronically coupled, that is, if the individual constituents are close enough for their wavefunctions to interact, allowing charge transport to occur. The properties of these films can be modified either by varying the properties of their individual constituents, such as their size and composition, or by tuning the strength of the electronic coupling, for example through ligand exchange in SQD films.[2, 3] Such bottom-up procedures for cheaply and easily constructing high-quality semiconductors from SQDs[4] and OSCs[5] have found applications in field effect transistors, photodetectors, solar cells and LEDs.

Au nanoclusters (NCs) also possess discrete energy levels, which, as in SQDs, are the result of quantum confinement arising from the small number of core Au atoms.[6, 7] Like OSCs, Au NCs are atomically precise, containing a specific number of core Au atoms, typically 100 or less. This makes them fundamentally different than SQDs, whose core size is defined by a size distribution in the nanometer range.[8] However, one key similarity between Au NCs and SQDs is that organic ligands surround their cores and make them colloidally stable. The core size and organic ligands determine the strength of electronic coupling in both SQD and Au NC films. For example, electrostatic Coulomb repulsion governs charge transport in highly resistive, electrically isolated films comprised of the extensively studied Au NCs with 55 core atoms and short aromatic ligands.[9] Increasing the electronic coupling by crosslinking the Au cores with dithiol ligands eliminates Coulombic effects and drastically increases the film's conductivity.[10] However, despite similarities to SQDs or OSCs, semiconducting behavior has not been observed in films of Au NCs of any size.

We report the observation of field effect and photoconductivity in phototransistors made from semiconducting films of $[Au_{25}(PPh_3)_{10}(SC_2H_4Ph)_5X_2]^{2+}$ NCs, where X=Cl or Br (abbreviated as $Au_{25}$). This work demonstrates the viability of metal NCs as a new class of low cost, solution processed semiconducting material that combines the atomic precision of OSCs with the possibility for ligand engineering and size/energy level tuning of SQDs. Precise control of these very characteristics was key to obtaining semiconducting $Au_{25}$ NC films, by preventing excessive electronic coupling and a concomitant transition into a metallic state. First, the number of core Au atoms was reduced to 25, from 55 or more in previous studies.[9, 10] Due to quantum confinement, this generates an essential increase in the energy gap between the highest occupied molecular orbital (HOMO) and lowest unoccupied molecular orbital (LUMO). Second, steric hindrance from the phenyl containing ligands controlled the distance between the Au cores, preventing their wavefunctions from interacting excessively.[4, 8] These characteristics limited the electronic coupling just enough to ensure that the film remained semiconducting. The importance of the core size and ligands was investigated by synthesizing and measuring the electronic properties of two other Au nanocluster systems that did not exhibit semiconducting behavior.

$Au_{25}$ NCs are synthesized in ambient at a maximum temperature of 40° C (see SI for details and Figure S1 for mass spectrometry). Thorough removal of synthesis byproducts is critical for obtaining stable, high quality transport in the films.[11] The $Au_{25}$ NC, shown in **Figure 1**a, is comprised of two back-to-back $Au_{13}$ sub-units sharing a common vertex,[12] with the longest dimension of the Au core being 1.1 nm. All of the Au atoms are bonded to organic ligands, except for the two end atoms, which are bonded to Cl or Br. The absorption spectrum of the purified $Au_{25}$ NC dispersion, shown in black in Fig. 1b, displays sharp, molecule-like optical transitions. The low energy peak at 674 nm corresponds to the HOMO-LUMO energy gap, whereas the high energy features below 500 nm correspond to HOMO-$n$ to LUMO+$n$ transitions, arising from transitions within the $Au_{13}$ sub-units.[7] The spectrum of a 16 nm thick (Figure S2) $Au_{25}$ NC film, spin coated on a quartz substrate, is also shown in Figure 1b. Its HOMO-LUMO peak has its maximum at 687 nm, corresponding to a red-shift of 40 meV (from 1.84 to 1.80 eV) compared to the solution, as shown in Figure 1c. Such red-shifting is indicative of electronic coupling between the individual NCs.[3, 13]

To probe their electronic properties, the NCs are integrated into a field effect transistor (FET) by spin coating films onto a substrate with interdigitated Au electrodes. The transfer and output curves in Figure 1d and 1e, respectively, demonstrate $n$-type field effect: clear semiconducting behavior in a film of metal NCs. A high ON/OFF ratio of $5 \times 10^4$ is obtained for a drain voltage $V_d$=6 V, and the charge carrier mobility approaches $10^{-5}$ cm$^2$V$^{-1}$s$^{-1}$ at $V_d$=20 V. The output curves (Figure 1e) reveal exponential behavior below a certain critical $V_d$. This Schottky-like behavior is due to an energy mismatch between the metal Au electrode Fermi energy and the electrical transport band in the semiconducting film formed from the LUMO orbitals of the Au NCs. By increasing $V_d$, electrons are able to overcome the charge injection energy barrier, and the current exhibits the expected transistor output. In the transistor OFF state, mobile ions[14] and charged byproducts in the film also play a role in charge transport (further details in SI and Figures S3 and S4).

Photoconductivity, shown in **Figure 2** and Figure S5, is further proof that the $Au_{25}$ NC films are semiconducting. Illumination with 635 nm light increases film conductivity and eliminates charge suppression at low $V_d$, as shown in Figure 2a. This is an indication of a photodiode effect, through which photoexcited charge carriers are able to overcome the energy barriers at the metal electrode/semiconducting film interface. Figure 2b shows that illumination shifts the device's transfer curve to lower voltages, effectively *n*-doping it and revealing a photogating effect.[15] This mechanism, illustrated in Figure 2c, requires that one type of charge carrier, in this case holes, have a much lower carrier mobility. Upon illumination, photoexcited holes will remain essentially trapped in comparison to the much higher mobility electrons. Depending on the device geometry and the material's mobility, the electrons can circulate through the device multiple times before recombining, producing a sensitivity enhancing gain mechanism. The low hole carrier mobility is confirmed by measuring the $Au_{25}$ NC film in ambient, where the films are exposed to electron accepting oxygen and water in the air, and therefore exhibit *p*-type behavior. Comparing the film's electrical characteristics in ambient, where it is *p*-type, and in vacuum, where it is *n*-type, allows us to compare the magnitude of the hole and electron mobility, respectively. As shown in Figure S6, the electron mobility is over $10^3$ times larger than the hole mobility. Trapping and/or scattering of hole charge carriers, interacting electrostatically with unbound $Br^-$ and $Cl^-$ ions in the film, could be responsible for the very low hole mobility. The device's much higher responsivity at weak incident power, shown in Figure 2d, is another indication of photogating.[15] The device's temporal response to light while in the ON state ($V_g$=60 V) is shown in Figure 2e. Through a photodiode effect, the energy mismatch at the contacts is likely responsible for the fast rise time of 3.6 ms when the transistor is in the ON state (gate voltage $V_g$=60 V). This response time is already comparable or better than optimized 2D material photodetectors.[16] The photocurrent fall time at high incident power (19.6 ms), and the overall response at low incident power is significantly slower. This slow, sensitive photodetection mechanism is a result of photogating, whose response is limited by the characteristic recombination lifetime of the photoexcited charge carriers.[15]

$[Au_{11}(PPh_3)_8Cl_2]Cl$[17] ($Au_{11}$) and $[Au_{25}(SC_2H_4Ph)_{18}]^{1-}$ ($Au_{25}$-$PET^{1-}$) were synthesized to test the effect of varying the ligands and number of core Au atoms, respectively, on the electrical properties of the Au NC films. Films from both of these clusters showed much lower electrical conductivity compared to $Au_{25}$, as shown in **Figure 3**a, and no semiconducting field effect or photoconduction. The $Au_{25}$-PET clusters can be readily oxidized by ambient oxygen, changing the charge state from -1 to 0.[18] Both of these $Au_{25}$-PET NCs have absorption spectra with features similar to the $Au_{25}$, as shown in Figure 3b. The $Au_{25}$-PET film (absorption spectra shown between that of the $Au_{25}$-$PET^0$ and $Au_{25}$-$PET^{1-}$ solution) was spin coated from the $Au_{25}$-$PET^{1-}$ solution, but lacks its characteristic peak at 444 nm. Indeed, the film's absorption spectrum appears to be similar to that of the charge-neutral $Au_{25}$-$PET^0$ NCs in solution, indicating that the film readily oxidizes in air and forms aggregates on the substrate surface, as shown in the dark field optical images in Figure S7. Additionally, in contrast to what was observed for $Au_{25}$, the HOMO-LUMO transition of the $Au_{25}$-PET is at the same location for both film and solution (around 680 nm), indicating that there is no electronic coupling between the individual $Au_{25}$-PET NCs. This, along with the film's chemical and structural instability, explains its poor electrical conductivity and lack of semiconducting properties.

In the $Au_{11}$ NCs, the lower number of core Au atoms results in stronger quantum confinement, and increases the energy of the HOMO-LUMO transition (418 nm, 2.97 eV) compared to the NCs with 25 core atoms (674 nm, 1.83 eV). Similar to the $Au_{25}$-PET film, the optical spectrum of the $Au_{11}$ film is not indicative of electronic coupling, which, along with the increased HOMO-LUMO gap, will impede charge transport, producing highly resistive, non-semiconducting films. The $Au_{25}$ NCs thus appear to be within a size/energy window that allows for an adequate amount of electronic coupling in the film: enough so that charges are able to flow through the film and collective semiconducting properties can emerge, but not so much that the films become metallic.

Having established that the $Au_{25}$ NC film is semiconducting and identified the reasons for this behavior, we proceeded to explore its charge transport mechanism, which was surprisingly similar to that of an amorphous OSC film. The curves in **Figure 4**a reveal that the charge carrier mobility $\mu$ increases with both the charge carrier density, proportional to $V_g$, and the applied electric field $F=V_d/L$, where $L$ is the device channel length. As shown in Figure 4b, at higher electric fields the mobility follows a Poole-Frenkel relationship $\log(\mu) \sim \beta F^{1/2}$,[19] where $\beta=1.2 \times 10^{-3}$ $(V/cm)^{1/2}$ is a fitting constant. Deviation from this behavior at low electric fields is due to the contact resistance of the device. In addition, the applied electric field strongly affects the mobility's temperature dependence, as shown in Figure 4c. The mobility's overall dependence on charge carrier density, electric field and its decrease at lower temperatures is consistent with a Gaussian disorder model (GDM). Often used to describe OSC films, the GDM assumes that charge transport occurs via thermally activated hopping between randomly distributed states. Moreover, the GDM assumes that the density of states (DOS) have a Gaussian energetic distribution of bandwidth $\sigma$.[20] At high temperatures, transport is dominated by upward hops from states located near the Fermi energy $\varepsilon_F$ to states near the so-called transport energy $\varepsilon_T$, as shown in the schematics of Figure 4a. As the carrier density, and thus $\varepsilon_F$, is increased with the gate voltage, it becomes easier for charge carriers to hop to the transport level, causing the observed mobility increase.[21] At very low temperatures, the temperature dependence is reduced because the lack of thermal energy forces charges to tunnel long distances to energetically similar states instead of to spatially closer, higher energy states.[20]

The flattening and eventual decrease of the mobility at high gate voltages, observed in Figure 4a, has been observed in highly doped, amorphous OSC films.[22] The unbound Br$^-$ and Cl$^-$ ions in the film may be acting as dopants and increasing the overall charge carrier concentration. The relatively low mobility (<10$^{-5}$ cm$^2$V$^{-1}$s$^{-1}$) may be a result of intrinsic (e.g. grain boundaries) and extrinsic (e.g. contacts) factors.[23]

The relationship between mobility, charge carrier concentration and temperature can be used to calculate the approximate width of the Gaussian DOS in the Au$_{25}$ NC film.[24] In the range where the mobility depends strongly on the charge carrier concentration (about 0<$V_g$<30 V), it follows an Arrhenius-like log($\mu$)~$T^{-1}$ dependence at high temperature, as shown in Figure 4d. The activation energy in this region decreases with increasing charge carrier concentration, as shown in Figure 4e. Extending the lines used to fit the activation energy shows that they intersect at a higher temperature. This behavior, which is further indication of the similarity to OSCs, can be used to calculate an approximate DOS width $\sigma=5k_BT_0/2=70$ meV, considering an effective medium approach within the GDM.[24] The bandwidth falls within the range of typical disordered, amorphous OSC films of small molecules or solution processed polymers.[24]

Semiconducting behavior has been demonstrated for the first time in a metal NC film. Charge transport could be improved in such films via ligand engineering, extrinsic doping, or even altering the atomic composition of the core. Further optimization could pave the way for exciting discoveries and device applications based on solution-processed, semiconducting metal NC films.

**Experimental Section**

*Materials:* The following chemicals were used as received. Triphenylphosphine (99%), ethanol (99.98%), acetonitrile (99,3%) and toluene (99.85%), were purchased from Acros Organics.

Methanol (99,8%), Tetrahydrofurane (min 99%), and n-hexane (96%) were purchased from VWR. Gold(III) chloride trihydrate (99,995%) and tetraoctylammonium bromide (98%) were obtained from Alfa Aesar. Pentane (99,9%) and dichlormethane (DCM, 99,5%) was obtained from Grüssing. Water was purified using a Millipore-Q System (18.2 MΩ cm).

*Au NC Synthesis*: $Au_{25}$ NCs were synthesized in two steps. First, polydisperse Au nanoparticles (1-3 nm diameter) are synthesized. Second, thiol etching of the Au nanoparticles was employed in order to obtain the desired $Au_{25}$ NCs with the formula $[Au_{25}(PPh_3)_{10}(PET)_5X_2]^{2+}$ (X= Cl or Br).[11] To start, $HAuCl_4 \cdot 3H_2O$ (0.100 g, dissolved in 3 mL of Millipore $H_2O$) was added into a 8 mL toluene solution of tetraoctylammonium bromide (TOAB, 0.145 g) and stirred vigorously (400 rpm) at room temperature for 15 min. When the aqueous phase became colorless and clear, indicating the complete transfer of the gold compound from aqueous to toluene phase, it was removed with a pipette. The organic phase was subsequently transferred using a glass pipette into the 3-neck flask. Subsequently, triphenylphosphine ($PPh_3$, 0.180 g) was added into the flask under stirring (800 rpm). Within few to tens of seconds, the solution became cloudy white. After 15 min, freshly dissolved sodium borhydride ($NaBH_4$, 0.026 g, dissolved by ultrasonication in 5 mL of EtOH) was injected rapidly to reduce $Au^I(PPH_3)X$ (starting material) to Au nanoparticles. After 2 hours, the black dispersion was dried by rotary evaporation at 50°C, resulting in a dry black solid. The black solid was mixed with 20 mL dicholormethane (DCM), vortexed and centrifuged at 16000 rpm for 3 min. The resulting black supernatant was then heated to 40°C under reflux. Phenylethanethiol (PET, 300 μL) was added to the black dispersion, which was stirred at 400 rpm. When the UV-Vis optical spectrum evolved to look like that in Fig. 1B, or latest after 96 h, the yellow/brownish dispersion was dried by rotary evaporation at 50°C. An oily black product was obtained, mixed with 2 mL DCM and transferred into a glass centrifuge flask. Subsequently, the suspension was precipitated with 80 mL of hexane, centrifuged at 3000 rpm and the supernatant was removed. The DCM/hexane washing step was repeated four more times ($V_{DCM}/V_{Hex}$= 1/20; 1/40; 0.5/40; 0.5/40). Finally, the $[Au_{25}(PPh_3)_{10}(SC_2H_4Ph)_5X_2]^{2+}$ were extracted by mixing with 10 mL MeOH, vortexing and centrifugation at 16000 rpm. $Au_{25}$-PET was synthesized according to procedures from Zhu et al[25] and Qian et al;[26] $Au_{11}$ according to McKenzie, et al.[17]

*Device Fabrication*: Field effect transistor devices were fabricated on highly *n*-doped Si wafers covered by 300 nm $SiO_2$. Two probe, interdigitated geometries were patterned on the wafer surface via optical lithography, thermal evaporation of 5/45 nm Ti/Au and liftoff. Au NC solutions (methanol for $Au_{25}$, tetrahydrofuran for $Au_{25}$-PET, dichloromethane for $Au_{11}$) were adjusted to have an optical density between 3 and 6 at 415 nm. Solutions were spin-coated onto the substrate with an acceleration of 250 rpm/s and speed of 4000 rpm/s for 20 s. The length *L* of the FET channel in these electrodes ranged from 5 to 15 µm. The device width *W* was either 99 channels by 600 µm long (59400 µm) or 79 channels by 500 µm long (39500 µm).

**Supporting Information**
Supporting Information is available from the Wiley Online Library or from the author.

**Acknowledgements**
We thank R. Schuster for mass spectrometry measurements. **Funding:** The authors gratefully acknowledge financial support of the European Research Council via the ERC Starting Grant "2D-SYNETRA" (Seventh Framework Programme FP7, Project: 304980). C. K. and M.G. thank the German Research Foundation DFG for financial support in the frame of the Cluster of Excellence "Center of ultrafast imaging CUI". C.K. thanks the Heisenberg scholarship KL 1453/9-2. M. Galchenko and A. Black contributed equally to this work.

**Figures**

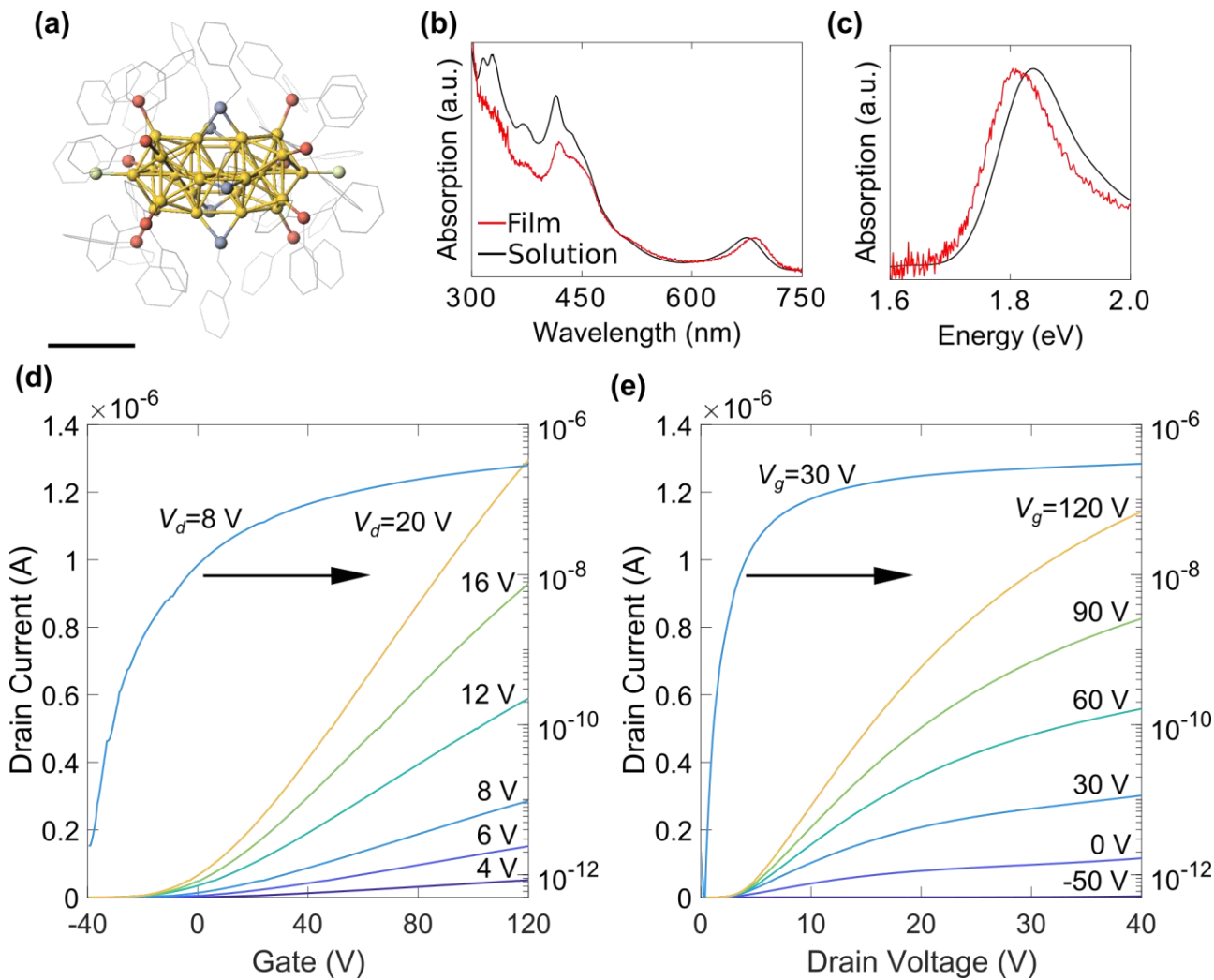

**Figure 1.** a) Schematic of $Au_{25}$ NC with yellow = Au, blue = P, red = S, green = Br or Cl. Scale bar 0.5 nm. Crystallographic data from Qian et al.[12] b) Optical absorption of $Au_{25}$ NC in solution and spin coated film. c) Closeup of optical absorption of HOMO-LUMO peak. d) Transfer and e) output curves of $Au_{25}$ film FET. Transfer $V_d$=8 V and output $V_g$=30 V curves shown also for logarithmic axes. $L$, $W$=10, 59400 μm.

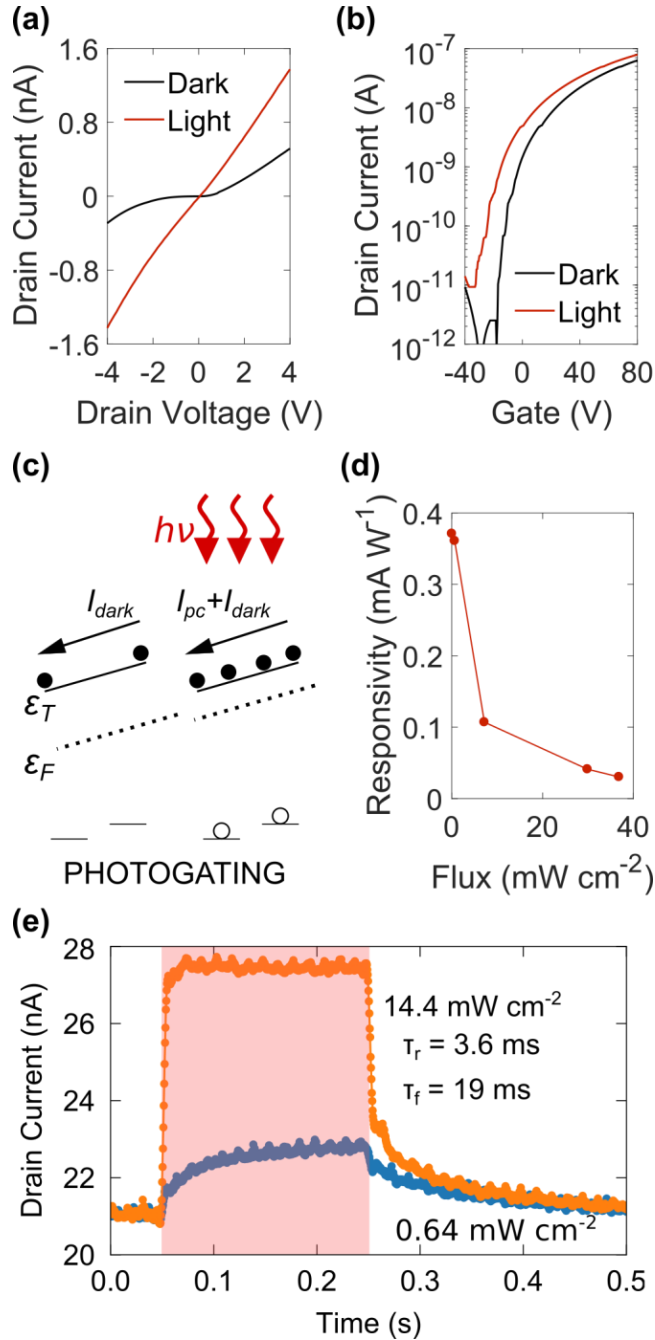

**Figure 2.** Device response using laser light with a wavelength of 635 nm. a) IV ($V_g$=0 V) and b) transfer ($V_d$=20 V) curve of dark and illuminated device. c) Schematics of illustrating photocurrent generation via photogating, showing Fermi energy $\varepsilon_F$ approaching transport energy $\varepsilon_T$ when illuminated. $I_{dark}$, $I_{pc}$ are dark and photocurrent, respectively. d) Device responsivity ($V_g$=0 V, $V_d$=4 V). e) Time response of photocurrent ($V_g$=0 V, $V_d$=4 V). $L, W$ = 10, 39500 μm.

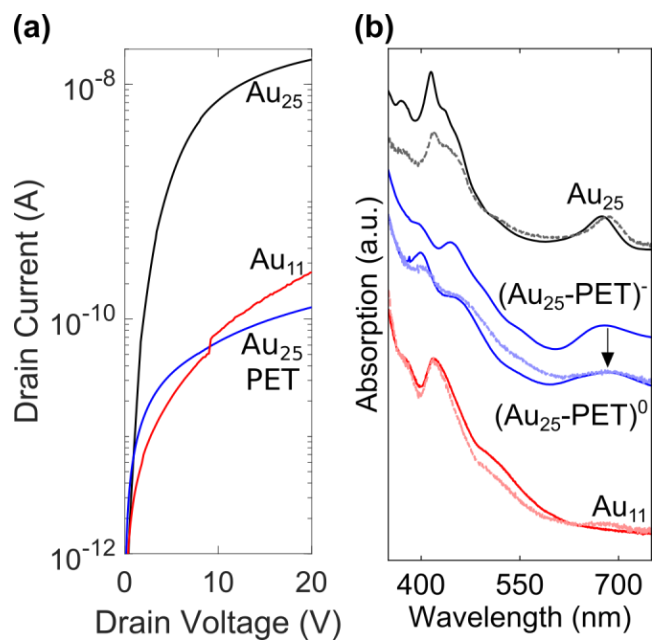

**Figure 3.** a) IV ($V_g$=0 V) curves of various Au nanocluter films. b) Optical absorption of various Au nanoclusters in solution (dark, solid lines) and films (dashed, light lines). Arrow indicates the oxidation of $Au_{25}$-$PET^{1-}$ to $Au_{25}$-$PET^0$ via $H_2O_2$ addition in solution or exposure to air in film. $L$, $W$ = 10, 39500 μm.

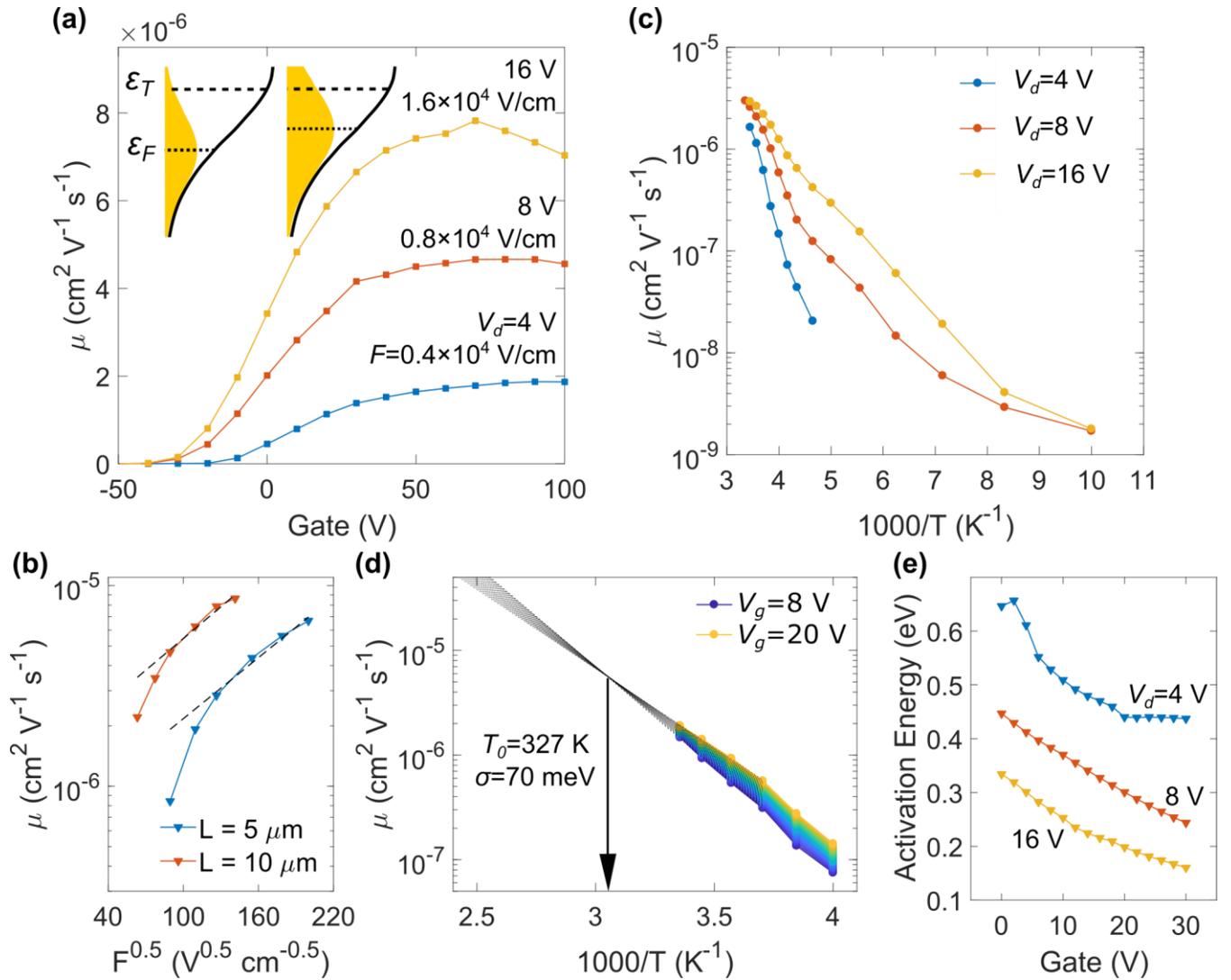

**Figure 4.** a) Mobility versus gate voltage curves at the indicated $V_{SD}$ and $F$. Schematic shows the Gaussian and occupied density of states (gold), along with the evolution of the Fermi and transport energies versus gate voltage. b) Mobility versus $F$ showing Poole-Frenkel behavior in the film. Arrhenius plot of the mobility over c) the entire temperature and d) at high temperature. e) Activation energies versus gate voltage. $L, W$ = 10, 39500 μm, except in b) where also $L$ = 5 μm.

# Supporting Information

**Field effect and photoconduction in Au$_{25}$ Nanoclusters Films**

*Michael Galchenko, Andrés Black, Leonard Heymann, and Christian Klinke\**

M. Galchenko,[+] Dr. A. Black,[+] L. Heymann, Prof. C. Klinke

Institute of Physical Chemistry, University of Hamburg, Martin-Luther-King Platz 6, 20146 Hamburg, Germany

E-mail: christian.klinke@swansea.ac.uk

Prof. C. Klinke

Department of Chemistry, Swansea University – Singleton Park, Swansea SA2 8PP, United Kingdom

Prof. C. Klinke

Institute of Physics, University of Rostock, 18059 Rostock, Germany

[+]These authors contributed equally to this work

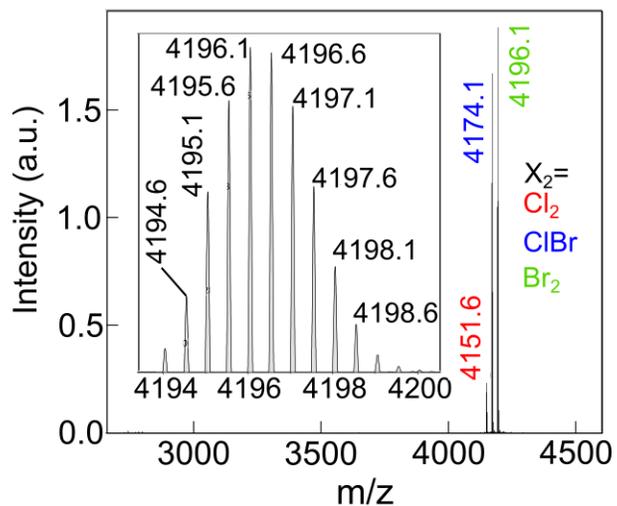

**Figure S1.** Mass spectrometry of the $[Au_{25}(PPh_3)_{10}(SC_2H_4Ph)_5X_2]^{2+}$ NC solution, showing X=Cl or Br. Inset: Closeup of $[Au_{25}(PPh_3)_{10}(SC_2H_4Ph)_5Br_2]^{2+}$ peak. The spacing of 0.5 in the isotope pattern indicates a total charge of $2^+$.

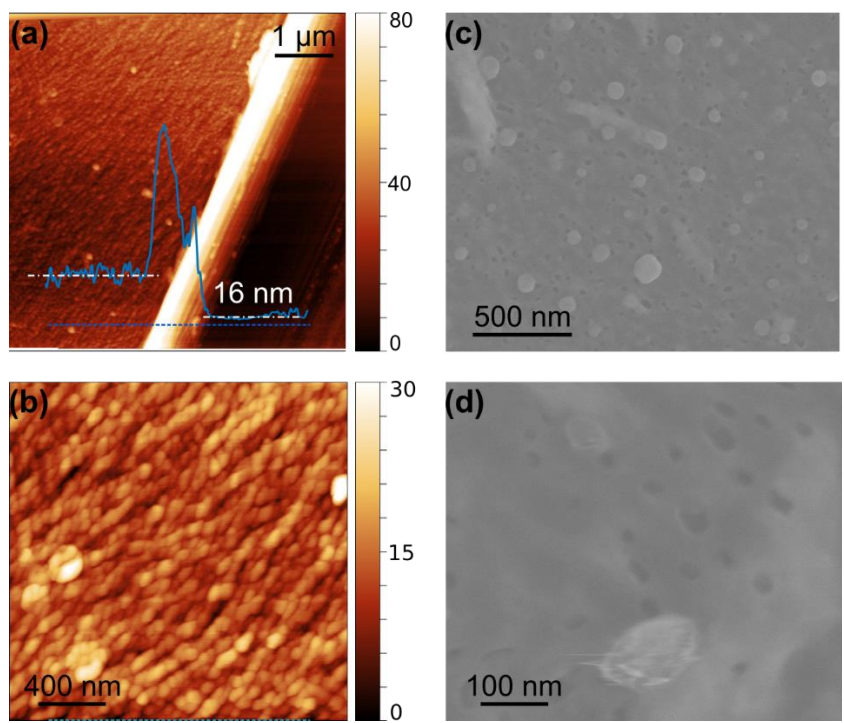

**Figure S2.** Au$_{25}$ NC spin-coated film morphology. a) and b) atomic force microscopy, revealing a film height of 16 nm and root-mean-square roughness of 3.2 nm. c) and d) scanning electron microscopy, revealing light-colored, round particles in the film about 100 nm in size. In addition, smaller, darker regions, which appear to be voids or holes in the film, are also observed. The particles and voids may result from impurities and drying effects due to the relatively fast film formation time (<20 s) from spin coating.

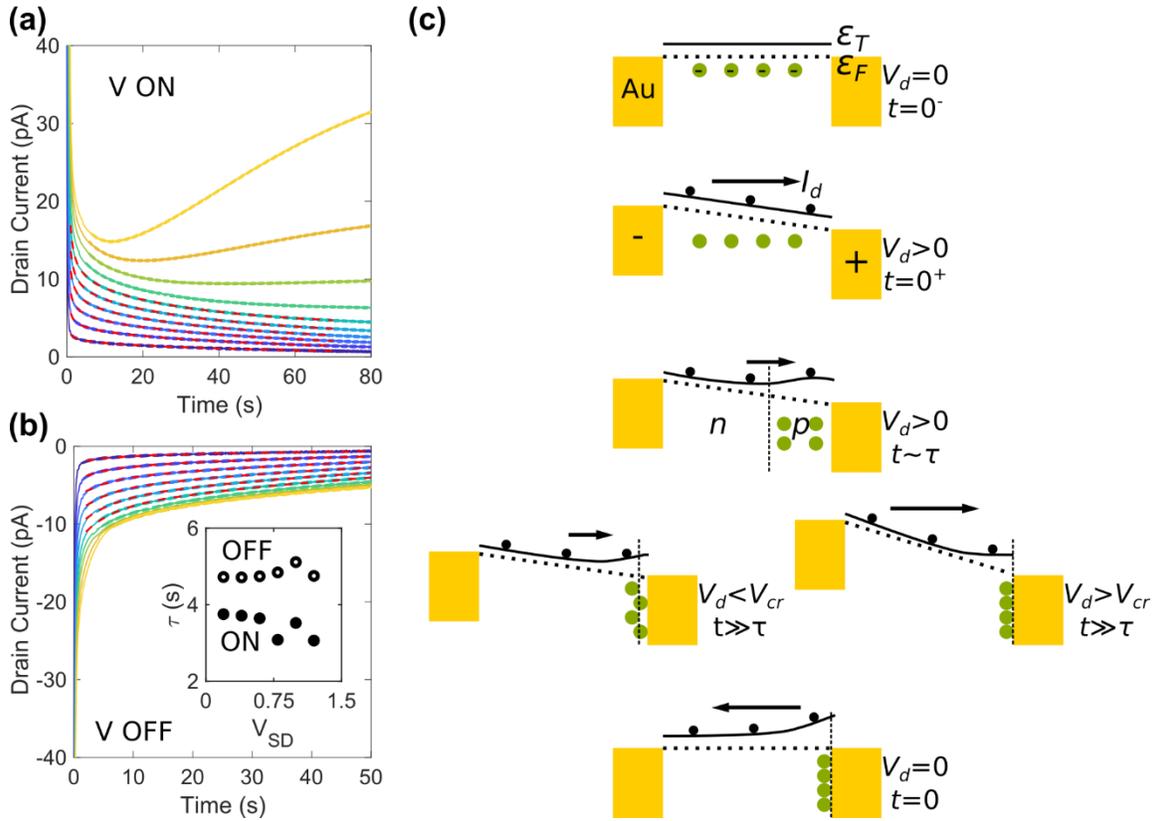

**Figure S3.** Ion motion in Au$_{25}$ NC films. Temporal response of source-drain current a) after applying source drain voltage. Curves go from $V_d$=0.2 V (blue) to 2 V (yellow) in 0.2 V interval. b) Temporal response of turning off the voltage ($V_d$=0) immediately following curves in a). Red dotted curves in a) and b) are fits. c) Schematic showing band bending due to ionic motion, ignoring Schottky contact effects for simplicity.

Figure S3a shows the temporal evolution of drain current $I_d$ after applying a constant drain voltage $V_d$ (V ON). For $V_d \leq 1.2$ V, the positive current decays continuously with time, eventually reaching a steady state value. In contrast, for $V_d > 1.2$ V, after an initial decay the current begins to increase. When the voltage is set back to 0, (V OFF, Fig. S3b), the current becomes negative, spiking before decaying towards 0. This behavior can be explained by ions, most likely unbound Br$^-$ or Cl$^-$ leftover from the synthesis, moving in response to the applied electric field.[1] The curves that experience pure decay ($V_d \leq 1.2$ V) are fit to a double exponential model $I_{SD}(t) = I_{ss} + Ae^{\tau_{eq}} + Be^{\tau}$ where $I_{ss}$ is the steady state current, $\tau_{eq}$ is the time response of the measuring equipment, $\tau$ is the time constant of the ions' motion in the film, and $A$ and $B$ are fitting constants.[2] The ions have a time constant between 3 and 4 seconds in the ON state, and between 4 and 5 seconds in the OFF state.

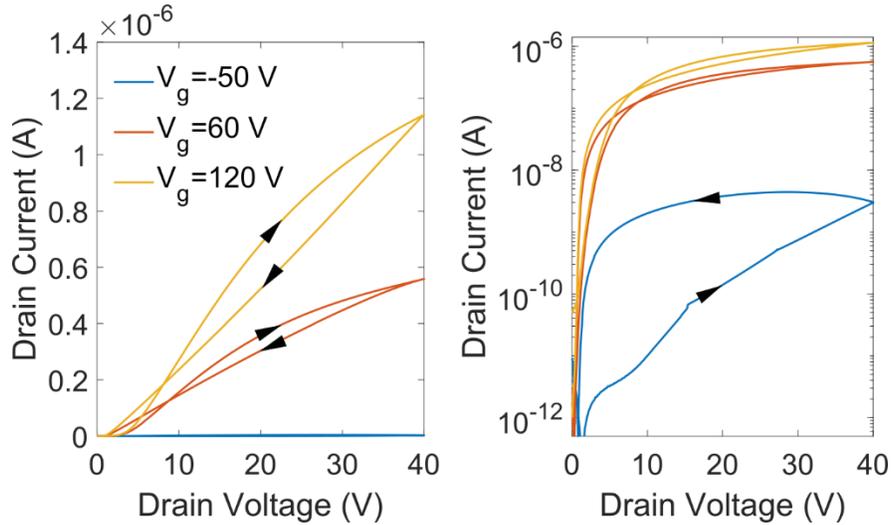

**Figure S4.** Forward and backwards output sweeps. $L$, $W$=10, 59400 µm.

A schematic of the ions' motion in response to the applied electric field is presented in Figure S3c. Initially ($t=0^-$), the mobile, negatively charged ions are assumed to be homogeneously distributed throughout the film. The moment a voltage is applied across the channel ($t=0^+$), a current is generated as electrons flow towards the positive electrode. The negatively charged ions also move (much more slowly than the electrons) towards the positive electrode, accumulating near it. The speed with which the ions drift towards the electrodes is reflected in the ionic motion time constant. In effect, for times $t\sim\tau$, a forward-biased *pn* junction forms in the device, with the accumulated negative ions corresponding to the *p* side. The ions accumulated near the electrode screen the applied electric field, reducing the current. For times much greater than $\tau$, the ion concentration in the vicinity of the positive constant is very high, effectively forming a $p^+n$ diode (a highly *p*-doped region). Higher applied electric fields can compress the ions into a smaller space, and thus achieve higher *p*-doping in the junction. Above a certain critical voltage $V_{cr}$, the ionic compression and applied electric field combine to achieve a flat band condition, eliminating the energy barrier and allowing current to flow freely. This is seen in Figure S3a for $V_d>V_{cr}$=1.2 V. After an initial decay, corresponding to a *pn* junction, the current increases as the $p^+n$ junction forms and the ions are increasingly compressed against the positive electrode. When the applied voltage is removed, the ions are still accumulated on one side of the channel, and generate an electric field with polarity opposite to the initially applied one. This results in the negative current spike seen in Figure S3b, with the decay corresponding to the ions diffusing back to their original homogeneous distribution. Hysteresis in the film's output curves, shown in Fig. S4, are likely a result of ionic motion. This hysteresis is especially large in the transistor OFF state, for $V_g$=-50 V.

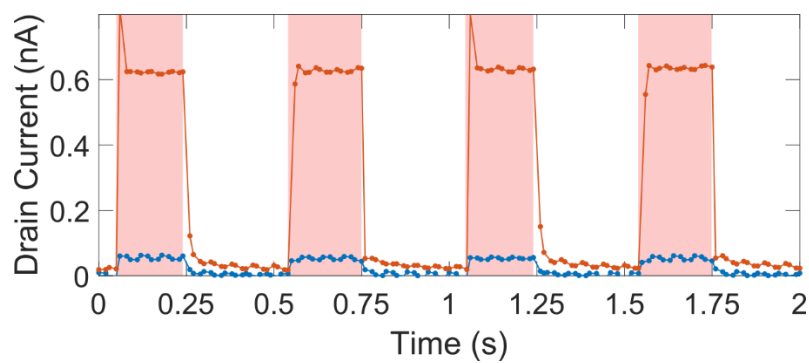

**Figure S5.** Photoresponse of device to 635 nm light for $V_d=4$ V, $V_g=0$ V. Blue and orange curves correspond to 0.64 and 29.9 mW cm$^{-2}$ incident power. $L$, $W$=10, 39500 μm.

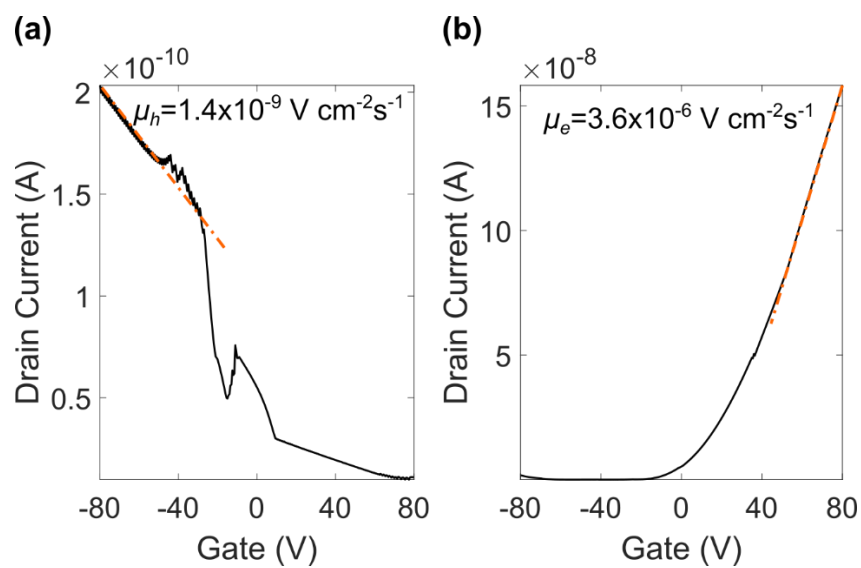

**Figure S6.** Transfer curve of a device in a) ambient and b) vacuum for $V_d=16$ V. $L$, $W$=10, 39500 μm.

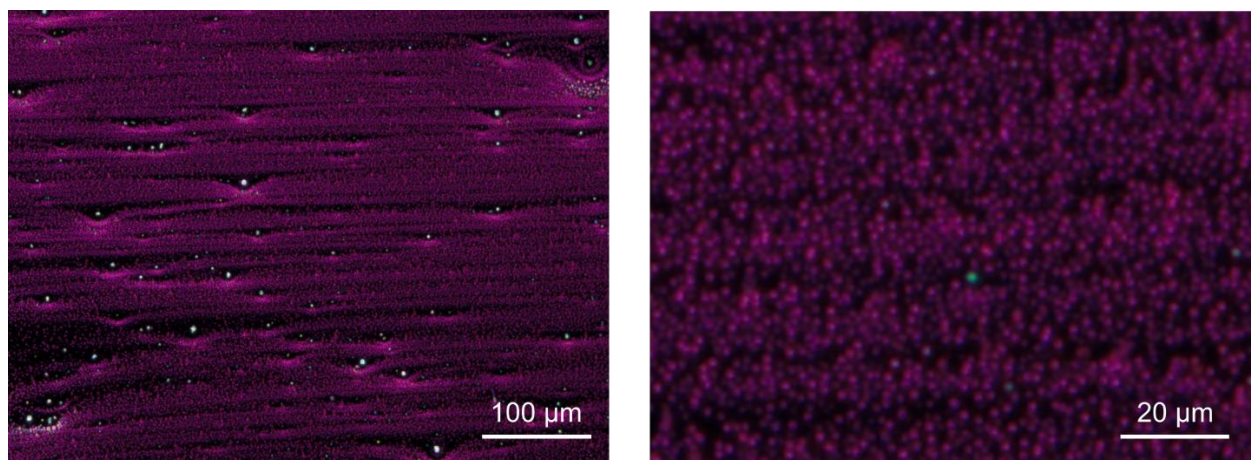

**Figure S7.** Dark field optical images of Au$_{25}$-PET film, showing aggregation of NCs.